# STUDY OF E-SMOOTH SUPPORT VECTOR REGRESSION AND COMPARISON WITH E- SUPPORT VECTOR REGRESSION AND POTENTIAL SUPPORT VECTOR MACHINES FOR PREDICTION FOR THE ANTITUBERCULAR ACTIVITY OF OXAZOLINES AND OXAZOLES DERIVATIVES

Doreswamy[1] and Chanabasayya .M. Vastrad[2]

[1]Department of Computer Science, MangaloreUniversity, Mangalagangotri-574 199, Karnataka,INDIA
Doreswamyh@yahoo.com

[2]Department of Computer Science, MangaloreUniversity, Mangalagangotri-574 199, Karnataka, INDIA
channu.vastrad@gmail.com

## ABSTRACT

*A new smoothing method for solving ε -support vector regression (ε-SVR), tolerating a small error in fitting a given data sets nonlinearly is proposed in this study. Which is a smooth unconstrained optimization reformulation of the traditional linear programming associated with a ε-insensitive support vector regression. We term this redeveloped problem as ε-smooth support vector regression (ε-SSVR). The performance and predictive ability of ε-SSVR are investigated and compared with other methods such as LIBSVM (ε-SVR) and P-SVM methods. In the present study, two Oxazolines and Oxazoles molecular descriptor data sets were evaluated. We demonstrate the merits of our algorithm in a series of experiments. Primary experimental results illustrate that our proposed approach improves the regression performance and the learning efficiency. In both studied cases, the predictive ability of the ε-SSVR model is comparable or superior to those obtained by LIBSVM and P-SVM. The results indicate that ε-SSVR can be used as an alternative powerful modeling method for regression studies. The experimental results show that the presented algorithm ε-SSVR, , plays better precisely and effectively than LIBSVMand P-SVM in predicting antitubercular activity.*

## 1.INTRODUCTION

The aim of this paper is supervised learning of real-valued functions. We study a sequence $S = \{(x_1, y_1), \ldots, (x_m, y_m)\}$of descriptor-target pairs, where the descriptors are vectors in $\mathbb{R}^n$ and the targets are real-valued scalars, $y_i \in \mathbb{R}$.Our aim is to learn a function $f: \mathbb{R}^n \to \mathbb{R}$ which serves a good closeness of the target values from their corresponding descriptor vectors. Such a function is usually mentioned to as a regression function or a regressor for short.The main aimof regression problems is to find a function $f(x)$that can rightly predict the target values,y of new input descriptor data points, x, by learning from the given training data set, S.Here, learning from a given training dataset means finding a linear surface that accepts a small error in fitting this training data set. Ignoring thevery smallerrors that fall within some acceptance, say εthat maylead to a improvedgeneralization ability is performed bymake use of an ε -insensitive loss function. As well as applying purpose of support vector machines (SVMs) [1-4], the function $f(x)$is made as flat as achievable, in fitting the training data. This issue is called ε -support vector regression (ε-SVR) and a descriptor data point$x^i \in R^n$is called a support vector if$\left|f(x^i) - y_i\right| \geq \varepsilon$.Generally, ε-SVR is developed as a constrained minimization problem [5-6], especially, a convex quadratic programming problem or a linear programming problem[7-9].Suchcreations presents 2m more nonnegative variablesand 2m inequality constraints that increase the problem sizeand could increase computational complexity for solvingthe problem. In our way, we change the model marginally and apply the smooth methods that have been widely used for solving important mathematical programming problems[10-14] and the support vector machine for classification[15]to deal with the problem as an unconstrained minimizationproblemstraightly. We name this reformulated problem as ε – smooth support vector regression(ε-SSVR). Because ofthe limit less arrangement of distinguishability of the objectivefunction of our unconstrained minimization problem, weuse a fast Newton-Armijo technique to deal with this reformulation. It has been shown that the sequence achieved by the Newton-Armijo technique combines to the unique solutionglobally and quadratically[15]. Taking benefit of ε-SSVR generation, we only need to solve a system of linear equations iteratively instead of solving a convex quadratic program or a linear program, as is the case with a conventionalε-SVR. Thus, we do not need to use anysophisticated optimization package tosolve ε-SSVR. In order to deal with the case of linear regression with aOxazolines and Oxazoles molecular descriptor dataset.

The proposed ε-SSVR model has strong mathematical properties, such as strong convexity and infinitely often differentiability. To demonstrate the proposed ε-SSVR's capability in solving regression problems, we employ ε-SSVR to predict ant tuberculosis activity for Oxazolines and Oxazoles agents. We also compared our ε-SSVR model with P-SVM[16-17] and LIBSVM [18] in the aspect of prediction accuracies. The proposed ε-SSVR algorithm is implemented in MATLAB.

A word about our representation and background material is given below. Entire vectors will be column vectors by way of this paper.For a vector xin the n-dimensional real descriptor space $R^n$ , the plus function$x_+$ is denoted as $(x_+)_i = \max\{0, x_i\}, i = 1, \ldots\ldots, n$. The scalar(inner) product of two vectors x and y in the n-dimensional real descriptor space $R^n$ will be reprsented by $x_T y$ and the p-norm of x will be represnted by $\|x\|_p$. For a matrix $A \in R^{m\times n}$, $A_i$ is the iTh row of A which is a row vector in$R^n$? A column vector of ones of arbitrary dimension will be reprsented by 1. For $A \in R^{m\times n}$ and $B \in R^{n\times l}$ , the kernel $K(A, B)$ maps $R^{m\times n} \times R^{n\times l}$ into$R^{m\times l}$. In exact, if x andy are column vectors in $R^n$ , then $K(x^T, y)$ is a real number , $K(A, x) = K(x^T, A^T)^T$ is a column vector in $R^m$. and$K(A, A^T)$ is an $m \times m$ matrix . If f is a real valued function interpreted on the n-dimensional real descriptor space$R^n$ , the gradient of f at x is represented by $\nabla f(x)$ which is a row vector in $R^n$ and $n \times n$ Hessian matrixof second partial derivatives of f at x is represented by$\nabla^2 f(x)$ . The base of the natural logarithm will be represented bye.

# 2. MATERIALS AND ALGORITHAMS

## 2.1 The Data Set

The molecular descriptors of 100Oxazolines and Oxazoles derivatives [19-20] based H37Rv inhibitors analyzed. These molecular descriptors are generated using Padel-Descriptor tool [21]. The dataset covers a diverse set of molecular descriptors with a wide range of inhibitory activities against H37Rv. The pIC50 (observed biological activity) values range from -1 to 3. The dataset can be arranged in data matrix. This data matrix $x$ contains $m$ samples (molecule structures) in rows and $n$ descriptors in columns. Vector $y$ with order $m \times 1$ denotes the measured activity of interest i.e. pIC50. Before modeling, the dataset is scaled.

## 2.2 The Smooth ε –support vector regression(ε-SSVR)

We allow a given dataset Swhich consists of m points in n-dimensional real descriptor space $R^n$ denoted by the matrix $A \in R^{m\times n}$ and m observations of real value associated with each descriptor. That is,$S = \{(A_i, y_i) | \; A_i \in R^n, y_i \in R, \text{for } i = 1, \ldots\ldots, m\}$we would like to search a nonlinear regression function,$f(x)$, accepting a small error in fitting this given data set. This can be performed by make use of the ε- insensitive loss function that sets ε- insensitive "tube" around the data, within which errors are rejected. Also, put into using the idea of support vector machines (SVMs) [1-4],thefunction f(x) is made as flat as possible in fitting thetraining data set. We start with the regression function f(x) and it is expressed as $f(x) = x^T w + b$. This problem can be formulated as an unconstrained minimization problem given as follows:

$$\min_{(w,b)\in R^{n+1}} \frac{1}{2} w^T w + C1^T |\xi|_\varepsilon \qquad (1)$$

Where $|\xi| \in R^m, (|\xi|_\varepsilon)_i = \max\{0, |A_i w + b + y_i| - \varepsilon\}$ that denotes the fitting errors and positive control parameter $C$ here weights the agreement between the fitting errors and the flatnessof the regression function$f(x)$. To handle ε-insensitive loss function in the objective function of the above minimization problem,traditionallyit is reformulated as a constrained minimization problem expressed as follows:

$$\min_{(w,b,\xi,\xi^*)} \frac{1}{2} w^T w + C1^T(\xi + \xi^*) Aw - 1b - y \leq 1\varepsilon + \xi - Aw - 1b + y \leq 1\varepsilon + \xi^* \; \xi, \xi^* \geq 0. \qquad (2)$$

This equation (2), which is equivalent to the equation (1), is a convex quadratic minimization problem with $n + 1$ free variables, $2m$ nonnegative variables, and $2m$ imparity constraints.However, presenting morevariables (and constraints in the formulation increases theproblem size and could increase computational complexityfor dealing with the regression problem.

In our smooth way, we change the model marginally and solve it as an unconstrained minimization problem preciselyapart from adding any new variable and constraint.

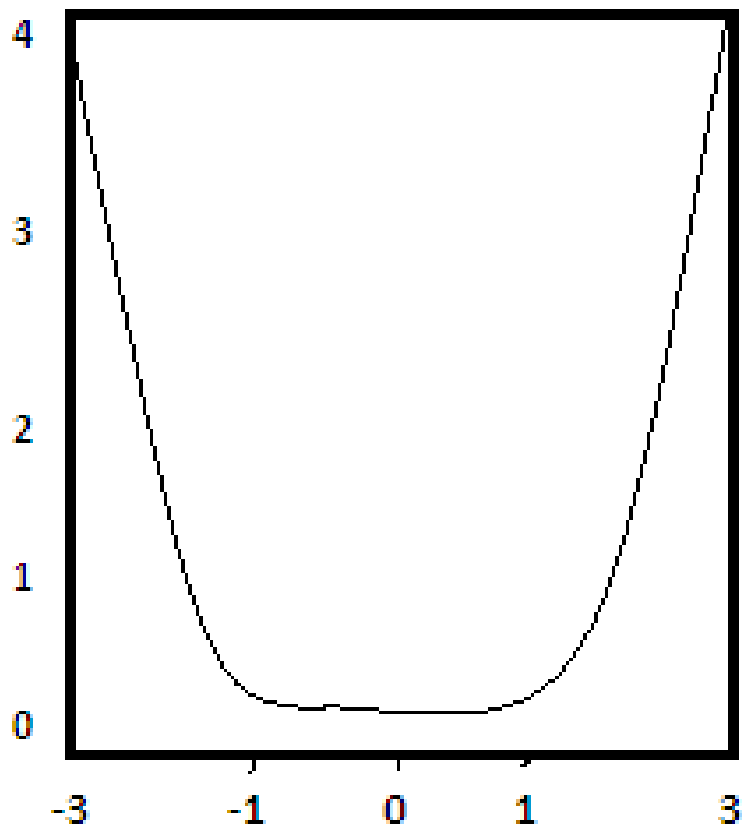

(a)

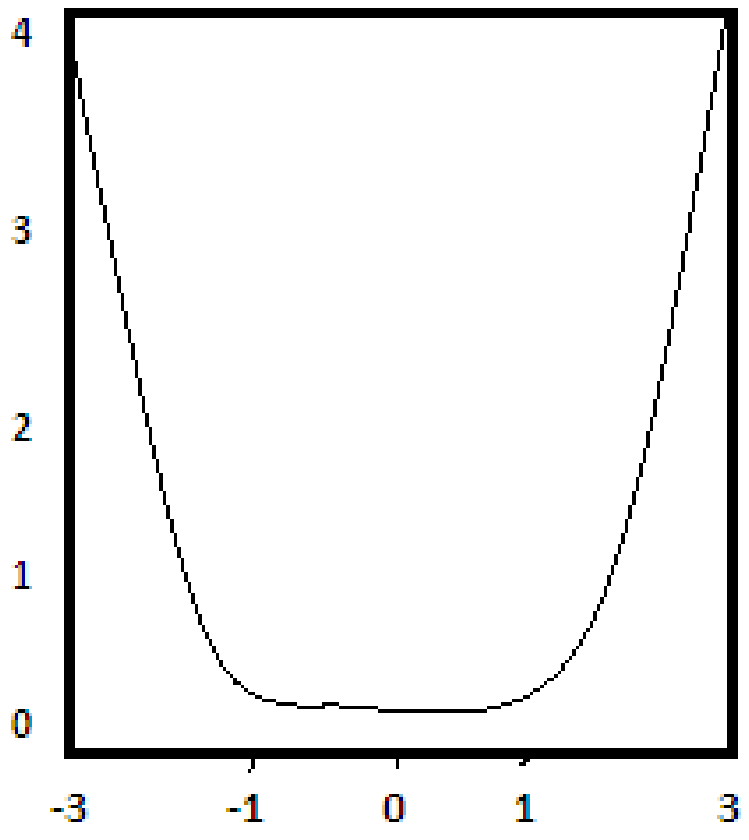

(b)

Figure1. (a) $|x|_\varepsilon^2$ and (b) $p_\varepsilon^2(x,\alpha)$ with $\alpha = 5, \varepsilon=1$.

That is, the squares of 2-norm ε- insensitive loss, $\| |Aw - 1b + y|_\varepsilon \|_2^2$ is minimized with weight $\frac{C}{2}$ in place of the 1-norm of ε- insensitive loss as in Eqn (1). Additional, we add the term $\frac{1}{2}b^2$ in the objective function to induce strong convexity and to certainty that the problem has a only global optimal solution. These produce the following unconstrained minimization problem:

$$\min_{(w,b)\in R^{n+1}} \frac{1}{2}(w^T w + b^2) + \frac{C}{2}\sum_{1=1}^{m} |A_i w + b - y_i|_\varepsilon^2 \qquad (3)$$

This formulation has been projected in active set support vector regression [22] and determined in its dual form. Motivated by smooth support vector machine for classification (SSVM) [15] the squares of ε- insensitive loss function in the above formulation can be correctly approximated by a smooth function which is extremely differentiable and described below. Thus, we are admitted to use a fast Newton-Armijo algorithm to determine the approximation problem. Before we make out the smooth approximation function, we exhibit some interesting observations:

$$|x|_\varepsilon = \max\{0, |x| - \varepsilon\}$$
$$= \max\{0, x - \varepsilon\} + \max\{0, -x - \varepsilon\} \qquad (4)$$
$$= (x-\varepsilon)_+ + (-x-\varepsilon)_+.$$

In addition, $(x-\varepsilon)_+ \, . \, (-x-\varepsilon)_+ = 0$ for all $x \in R$ and $\varepsilon > 0$. Thus, we have

$$|x|_\varepsilon^2 = (x-\varepsilon)_+^2 + (-x-\varepsilon)_+^2. \qquad (5)$$

In SSVM [15], the plus function $x_+$ approximated by a smooth p-function, $p(x,\alpha) = x + \frac{1}{\alpha}\log(1 + e^{-\alpha x}), \alpha > 0$. It is straightforward to put in place of $|x|_\varepsilon^2$ by a very correct smooth approximation is given by:

$$p_\varepsilon^2(x,\alpha) = \left(p(x-\varepsilon,\alpha)\right)^2 + \left(p(-x-\varepsilon,\alpha)\right)^2. \qquad (6)$$

Figure1.exemplifies the square of ε- insensitive loss function and its smooth approximation in the case of $\alpha = 5$ and $\varepsilon = 1$. We call this approximation $p_\varepsilon^2$-function with smoothing parameterα. This $p_\varepsilon^2$-function is used hereto put in place of the squares of ε- insensitive loss function of Eqn. (3) to get our smooth support vector regression(ε-SSVR):

$$\min_{(w,b)\in R^{n+1}} \Phi_{\varepsilon,\alpha}(w,b)$$
$$:= \min_{(w,b)\in R^{n+1}} \frac{1}{2}\left(w^T w + b^2\right)$$
$$+\frac{C}{2}\sum_{i=1}^{m} p_\varepsilon^2\,(A_i w + b - y_i, \alpha) \qquad (7)$$
$$= \min_{(w,b)\in R^{n+1}} \frac{1}{2}\left(w^T w + b^2\right)$$
$$+\frac{C}{2} 1^T p_\varepsilon^2(A_i w + b - y, \alpha),$$

Where $p_\varepsilon^2(A_i w + b - y, \alpha) \in R^m$ is expressed by $p_\varepsilon^2(A_i w + b - y, \alpha)_i = p_\varepsilon^2(A_i w + b - y_i, \alpha)$.

This problem is a powerfully convex minimization problem without any restriction. It is not difficult to show that it has a one and only solution. Additionally, the objective function in Eqn. (7)is extremelydifferentiable, thus we can use a fast Newton-Armijo technique to deal with the problem.

Before we deal with the problem in Eqn. (7) we have to show that the result of the equation (3) can be got by analyzing Eqn. (7) with α nearing infinity.

We begin with a simple heading thatlimits the difference betweenthe squares of ε- insensitive loss function,$|x|_\varepsilon^2$ and its smooth approximation $p_\varepsilon^2(x, \alpha)$.

***Heading 2.2.1.***For $x \in R$and $|x| < \sigma + \varepsilon$:

$$p_\varepsilon^2(x, \alpha) - |x|_\varepsilon^2 \le 2\left(\frac{\log 2}{\alpha}\right)^2 + \frac{2\sigma}{\alpha}\log 2, \qquad (8)$$

where $p_\varepsilon^2(x, \alpha)$is expressed in Eqn. (6).

***Proof.*** We allow for three cases. For $-\varepsilon \le x \le \varepsilon, |x|_\varepsilon = 0$ and $p(x, \alpha)^2$ are a continuity increasing function, so we have

$$p_\varepsilon^2(x, \alpha) - |x|_\varepsilon^2 = p(x - \varepsilon, \alpha)^2 + p(-x - \varepsilon, \alpha)^2$$
$$\le 2p(0, \alpha)^2 = 2\left(\frac{\log 2}{\alpha}\right)^2,$$

since$x - \varepsilon \le 0$ and $-x - \varepsilon \le 0$.

For $\varepsilon < x < \varepsilon + \sigma$, using the result in SSVM[15] that $p(x, \alpha)^2 - (x_+)^2 \le \left(\frac{\log 2}{\alpha}\right)^2 + \frac{2\sigma}{\alpha}\log 2$ for $|x| < \sigma$, we have

$$p_\varepsilon^2(x, \alpha) - (|x|_\varepsilon)^2$$
$$= (p(x - \varepsilon, \alpha))^2 + (p(-x - \varepsilon, \alpha))^2 - (x - \varepsilon)_+^2$$

$$\leq \left(p(x-\varepsilon,\alpha)\right)^2 - (x-\varepsilon)_+^2 + (p(0,\alpha))^2$$
$$\leq 2\left(\frac{\log 2}{\alpha}\right)^2 + \frac{2\sigma}{\alpha}\log 2.$$

Likewise, for the case of $-\varepsilon - \sigma < x < -\varepsilon$, we have

$$p_\varepsilon^2(x,\alpha) - (|x|_\varepsilon)^2 \leq 2\left(\frac{\log 2}{\alpha}\right)^2 + \frac{2\sigma}{\alpha}\log 2.$$

Hence, $p_\varepsilon^2(x,\alpha) - |x|_\varepsilon^2 \leq 2\left(\frac{\log 2}{\alpha}\right)^2 + \frac{2\sigma}{\alpha}\log 2.$

By Heading 2.1, we have that as the smoothing parameter α reaches infinity, the one and only solution of Equation (7) reaches, the one and only solution of Equation (3). We shall do this for a function $f_\varepsilon(x)$ given in Eqn. (9) below that includes the objective function of Eqn. (3) and for a function $g_\varepsilon(x,\alpha)$ given in Eqn. (10) below which includes the SSVR function of Eqn. (7).

***Axiom 2.2.2***. Let $A \in R^{m\times n}$ andb $\in R^{m\times 1}$. Explain the real valued functions $f_\varepsilon(x)$ and $g_\varepsilon(x,\alpha)$ in the n-dimensional real molecular descriptor space$R^n$:

$$f_\varepsilon(x) = \frac{1}{2}\sum_{i=1}^{m}\left|A_j x - b\right|_\varepsilon^2 + \frac{1}{2}\|x\|_2^2 \tag{9}$$

And

$$g_\varepsilon(x,\alpha) = \frac{1}{2}\sum_{i=1}^{m} p_\varepsilon^2(A_j x - b,\alpha) + \frac{1}{2}\|x\|_2^2, \tag{10}$$

With$\varepsilon,\alpha > 0$.

1. There exists a one and only solution $\bar{x}$ of $\min_{x\in R^n} f_\varepsilon(x)$ and one and only solution $\bar{x}_\alpha$of $\min_{x\in R^n} g_\varepsilon(x,\alpha)$.

2. For all $\alpha > 0$, we have the following inequality:

$$\|\bar{x}_\alpha - \bar{x}\|_2^2 \leq m\left(\left(\frac{\log 2}{\alpha}\right)^2 + \xi\frac{\log 2}{\alpha}\right), \tag{11}$$

Where$\xi$ is expressed as follows:

$$\xi = \max_{1\leq i\leq m} |(A\bar{x} - b)_i|. \tag{12}$$

Thus $\bar{x}_\alpha$ gathers to $\bar{x}$as α goes to endlessness with an upper limit given by Eqn. (11).

The proof can be adapted from the results in SSVM [15] and, thus, excluded here. We now express a Newton-Armijo algorithm for solving the smooth equation (7).

### 2.2.1A NEWTON-ARMIJO ALGORITHM FOR ε-SSVR

By utilizing the results of the preceding section and taking benefitof the twice differentiability of the objectivefunction in Eqn. (7), we determine a globally and quadratically convergent Newton-Armijo algorithm for solving Eqn. (7).

***Algorithm 2.3.1 Newton-ArmijoAlgorithm For ε-SSVR***

Start with any choice of initial point $(w^0, b_0) \in R^{n+1}$. Having $(w^i, b_i)$, terminate if the gradient of the objective function of Eqn. (7) is zero, that is, $\nabla\Phi_{\varepsilon,\alpha}(w^i, b_i)=0$. Else calculate $(w^{i+1}, b_{i+1})$ as follows:

1. ***Newton Direction:*** Decide the direction $d^i \in R^{n+1}$ by allocatingequal to zero the Linearization of $\nabla\Phi_{\varepsilon,\alpha}(w, b)$ all over $(w^i, b_i)$, which results in $n + 1$ Linear equations with $n + 1$ variables:

$$\nabla^2\Phi_{\varepsilon,\alpha}(w^i, b_i)d^i = -\nabla\Phi_{\varepsilon,\alpha}(w^i, b_i)^T. \quad (13)$$

2. ***Armijo Step size [1]:*** Choose a stepsize $\lambda_i \in R$ such that:

$$\left(w^{i+1}, b_{i+1}\right) = (w^i, b_i)+\lambda_i d^i , \quad (14)$$

where $\lambda_i = \max\{1, \frac{1}{2}, \frac{1}{4}, \ldots\ldots\}$ such that:

$$\Phi_{\varepsilon,\alpha}\left(w^i, b_i\right) - \Phi_{\varepsilon,\alpha}((w^i, b_i) +\lambda_i d^i \geq -\delta\lambda_i\Phi_{\varepsilon,\alpha}(w^i, b_i)d^i, \quad (15)$$

where $\delta \in \left(0, \frac{1}{2}\right)$.

Note that animportant difference between our smoothingapproach and that of the traditional SVR [7-9] is that we are solving a linear system of equations (13) here, rather solving a quadratic program, as is the case with the conventional SVR.

## 2.3LIBSVM

LIBSVM [18] is a library for support vector machines. LIBSVM is currentlyone of the most widely used SVM software. This software contains C-support vector classification (C-SVC), v-support vector classification (v-SVC), ε-support vector regression (ε-SVR), v-support vector regression (v-SVR). All SVM formulations supported in LIBSVM are quadratic minimization problems□

## 2.4Potential-Support Vector Machines(P-SVM)

P-SVM [16-17] is a supervised learning method used for classification and regression. As well as standard Support Vector Machines, it is based on kernels. Kernel Methods approach the problem by mapping the data into a high dimensional feature space, where each coordinate corresponds to one feature of the data items, transforming the data into a set of points in a Euclidean space. In that space, a variety of methods can be used to find relations between the data.

## 2.5Experimental Evaluation

In order to evaluate how well each method generalizedto unseen data, we split the entire data set into two parts,the training set and testing set. The training data was usedto generate the regression function that is learning fromtraining data; the testing set, which is not involved in thetraining procedure, was used to evaluate the predictionability of the resulting regression function.We also used a tabular structure scheme in splitting the entire data set to keep the "similarity" between training and testing data sets [23]. That is, we tried to make the training set and the testing set have the similar observation distributions. A smaller testing error indicates better prediction ability. We performed tenfold cross-validation on each data set [24] and reported the average testing error in our numerical results. Table 1 gives features of two descriptor datasets.

Table 1: Features of two descriptor datasets

| Data set(Molecular Descriptors of Oxazolines and Oxazoles Derivatives) | Train Size | Test Size | Attributes |
|---|---|---|---|
| Full | 75 X 254 | 25 X 254 | 254 |
| Reduced | 75 X 71 | 25 X 71 | 71 |

In all experiments, 2-norm relative error was chosen to evaluate the tolerance between the predicted values and the observations. For an observation vector y and the predicted vector $\hat{y}$ , the 2-norm relative error (SRE) of two vectors y and $\hat{y}$ was defined as follows.

$$\mathrm{SRE} = \frac{\|y - \hat{y}\|_2}{\|y\|_2} \qquad (16)$$

In statistics, the mean absolute error is a quantity used to measure how close predictions are to the eventual outcomes. The mean absolute error (MAE) is given by

$$\mathrm{MAE} = \frac{1}{n}\sum_{i=1}^{n}|\hat{y}_i - y_i| = \frac{1}{n}\sum_{i=1}^{n}|e_i| \qquad (17)$$

As the name suggests, the mean absolute error is an average of the absolute errors $e_i = \hat{y}_i - y_i$, where $\hat{y}_i$ is the prediction and $y_i$ the observed value.

In statistics, the coefficient of determination, denoted $R^2$ and pronounced R squared, is used in the context of statistical models whose main purpose is the prediction of future outcomes on the basis of other related information. $R^2$is most often seen as a number between 0 and 1, used to describe how well a regression line fits a set of data. A $R^2$ near 1 indicates that a regression line fits the data well, while a $R^2$ close to 0 indicates a regression line does not fit the data very well. It is the proportion of variability in a data set that is accounted for by the statistical model. It provides a measure of how well future outcomes are likely to be predicted by the model.

$$R^2 = 1 - \frac{\sum_{i=1}^{n}(y_i - \hat{y}_i)^2}{\sum_{i=1}^{n}(y_i - \bar{y})^2} \qquad (18)$$

The predictive power of the models developed on the calculated statistical parameters standard error of prediction (SEP) and relative error of prediction (REP)% as follows:

$$SEP = \left[\frac{\sum_{i=1}^{n}(\hat{y}_i - y_i)^2}{n}\right]^{0.5} \qquad (19)$$

$$REP(\%) = \frac{100}{\bar{y}}\left[\frac{1}{n}\sum_{i=1}^{n}(\hat{y}_i - y_i)^2\right]^{0.5} \qquad (20)$$

The performancesof models were evaluated in terms of root mean square error (RMSE), which was defined as below:

$$RMSE = \sqrt{\frac{\sum_{i=1}^{n}\left(y_i - \hat{y}_i\right)^2}{n}} \qquad (21)$$

Where$\hat{y}_i$ ,$y_i$and $\bar{y}$are the predicted, observed and mean activity property, respectively.

## 3.RESULTS AND DISCUSSION

In this section, we demonstrate the effectiveness of our proposed approachε-SSVR by comparing it to LIBSVM (ε-SVR) and P-SVM. In the following experiments, training is done with Gaussian kernel function $k(x1, x2) = \exp\left(-\Upsilon\|x_i - x_j\|^2\right)$, where $\Upsilon$is the is the width of the Gaussian kernel, $i, j = 1, \ldots\ldots, l$. We perform tenfold cross-validation on each dataset and record the average testing error in our numerical results. The performances of ε-SSVR for regression depend on the combination of several parameters They are capacity parameter $C$, ε of ε- insensitive loss function and $\Upsilon$parameter. $C$ is a regularization parameter that controls the tradeoff between maximizing the margin and minimizing the training error. In practice the parameter $C$ is varied through a wide range of values and the optimal performance assessed using a separate test set. Regularization parameter $C$, whose effect on the RMSE is shown in Figure 1a for full descriptor datasetandFigure 1b for reduced descriptor dataset.

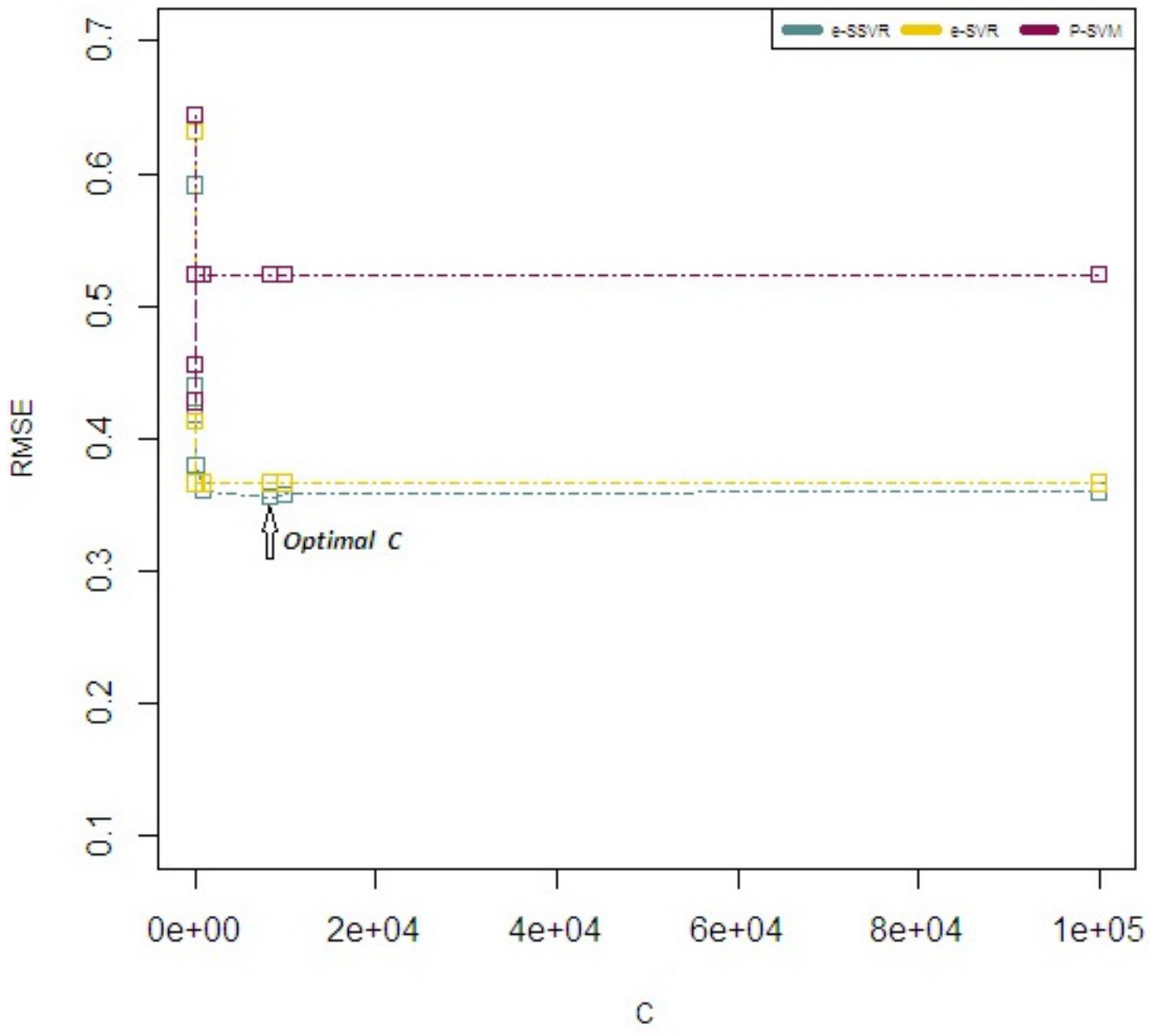

Figure 1a.The selection of the optimal capacity factor$C$ (8350) for ε-SSVR(ε=0.1, ϒ=0.0217)

For the Full descriptordataset, The RMSE valuefor ε-SSVRmodel 0.3563 is small for selected optimal parameter C, compared to RMSE values for other two models i.e. LIBSVM (ε-SVR) and P-SVM are 0.3665 and 0.5237.Similarly,for the reduced descriptor dataset,The RMSE value for ε-SSVR model 0.3339 is small for selected optimal parameter C, compared to RMSE values for other two models i.e. LIBSVM (ε-SVR) and P-SVM are 0.3791 and 0.5237. The optimal value for ε depends on the type of noise present in the data, which is usually unknown. Even if enough knowledge of the noise is available to select an optimal value for ε, there is the practical consideration of the number of resulting support vectors. E insensitivity prevents the entire training set meeting boundary conditions and so allows for the possibility of sparsely in the dual formulations solution.

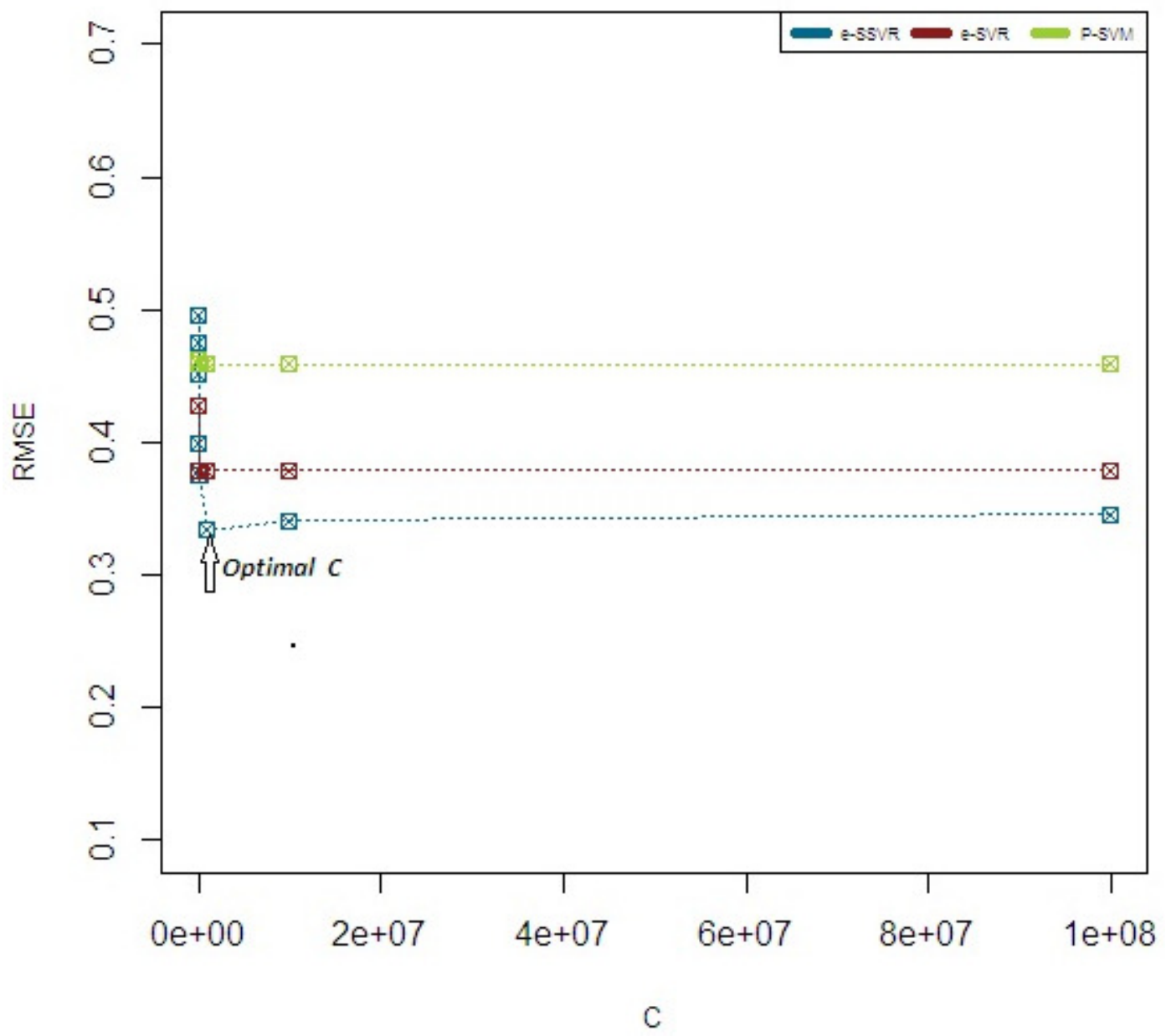

Figure 1b. The selection of the optimal capacity factor$C$(1000000) for ε-SSVR(ε=0.1, ϒ=0.02)

So, choosing the appropriate value of ε is critical from theory. To find an optimal ε, the root mean squares error (RMSE) on LOO cross-validation on different ε was calculated. The curves of RMSE versus the epsilon (ε) is shown in Figure 2a and Figure 2b.

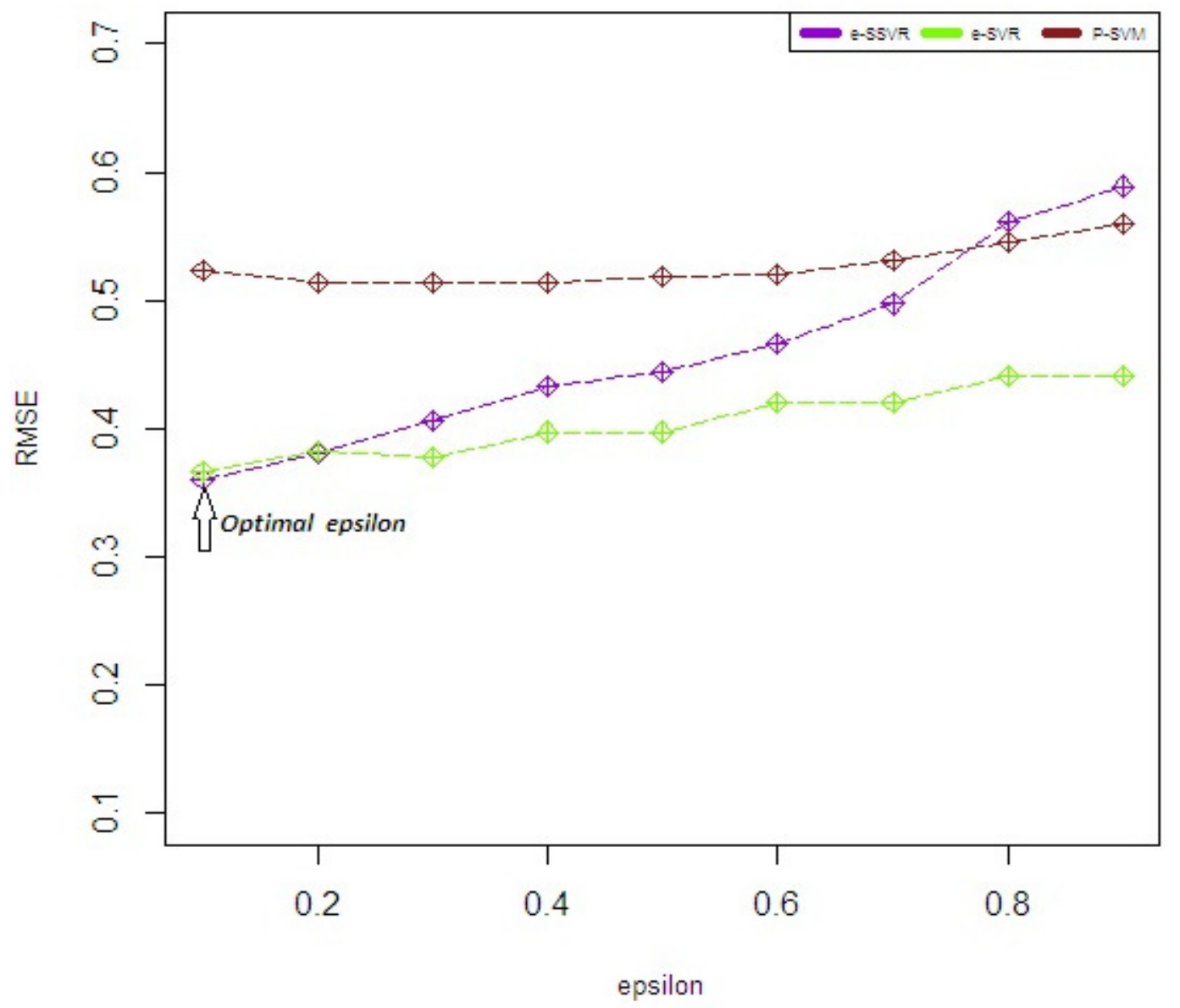

Figure 2a. The selection of the optimal epsilon (0.1) for ε-SSVR($C$ = 1000, ϒ=0.02)

For the Full descriptor dataset , The RMSE value for ε-SSVR model 0.3605 is small for selected optimal epsilon(ε), compared to RMSE value for LIBSVM(ε-SVR) model is closer i.e. 0.3665 but comparable to the proposed model and bigRMSE value for P-SVM model is 0.5237.

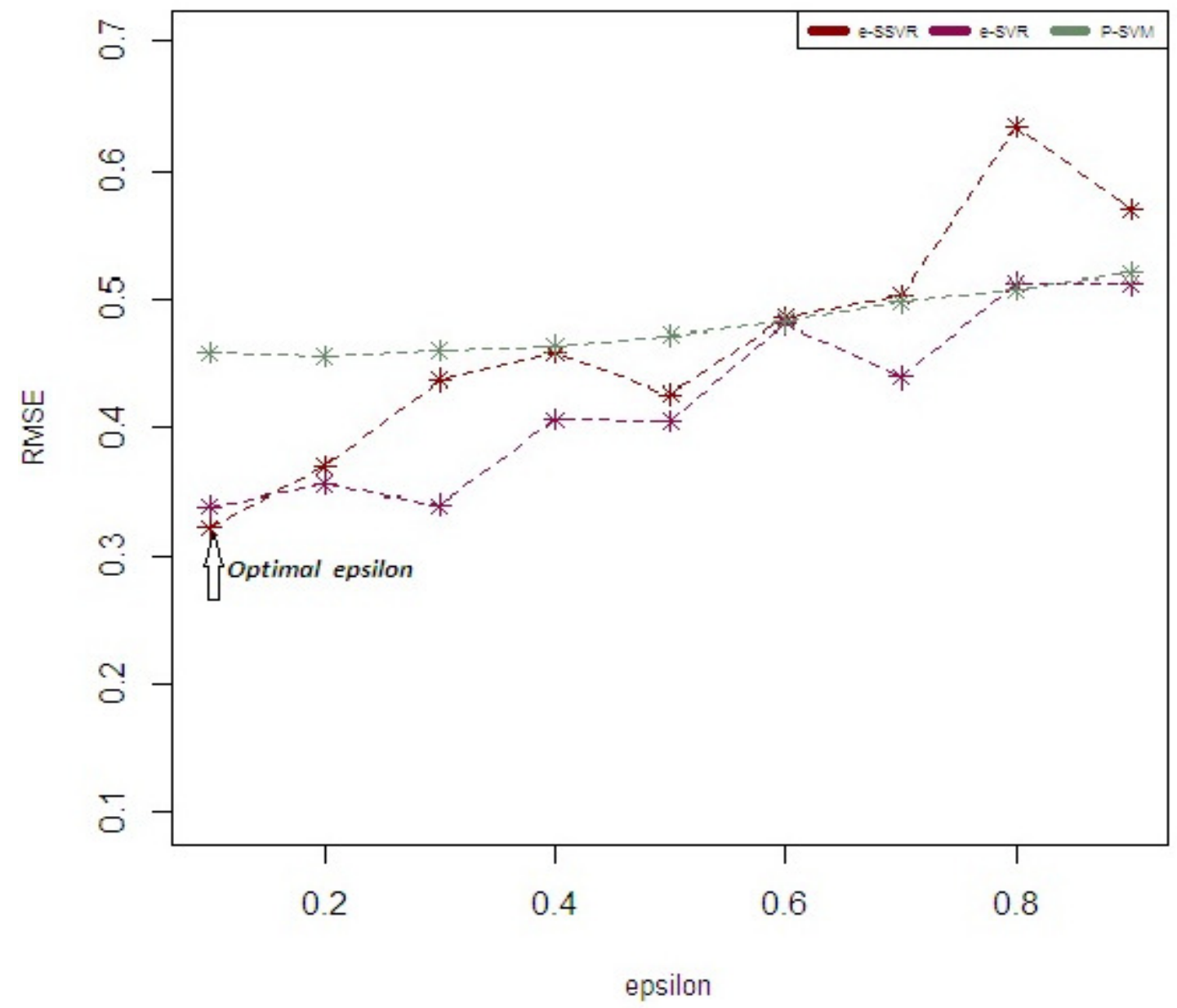

Figure 2b.The selection of the optimal epsilon (0.1) for ε-SSVR($C$= 10000000, ϒ=0.01)

Similarly , for the Reduced descriptor dataset , The RMSE value for ε-SSVR model 0.3216 is small for selected optimal epsilon(ε) , compared to RMSE values for other two models i.e. LIBSVM(ε-SVR) and P-SVM are 0.3386 and 0.4579.

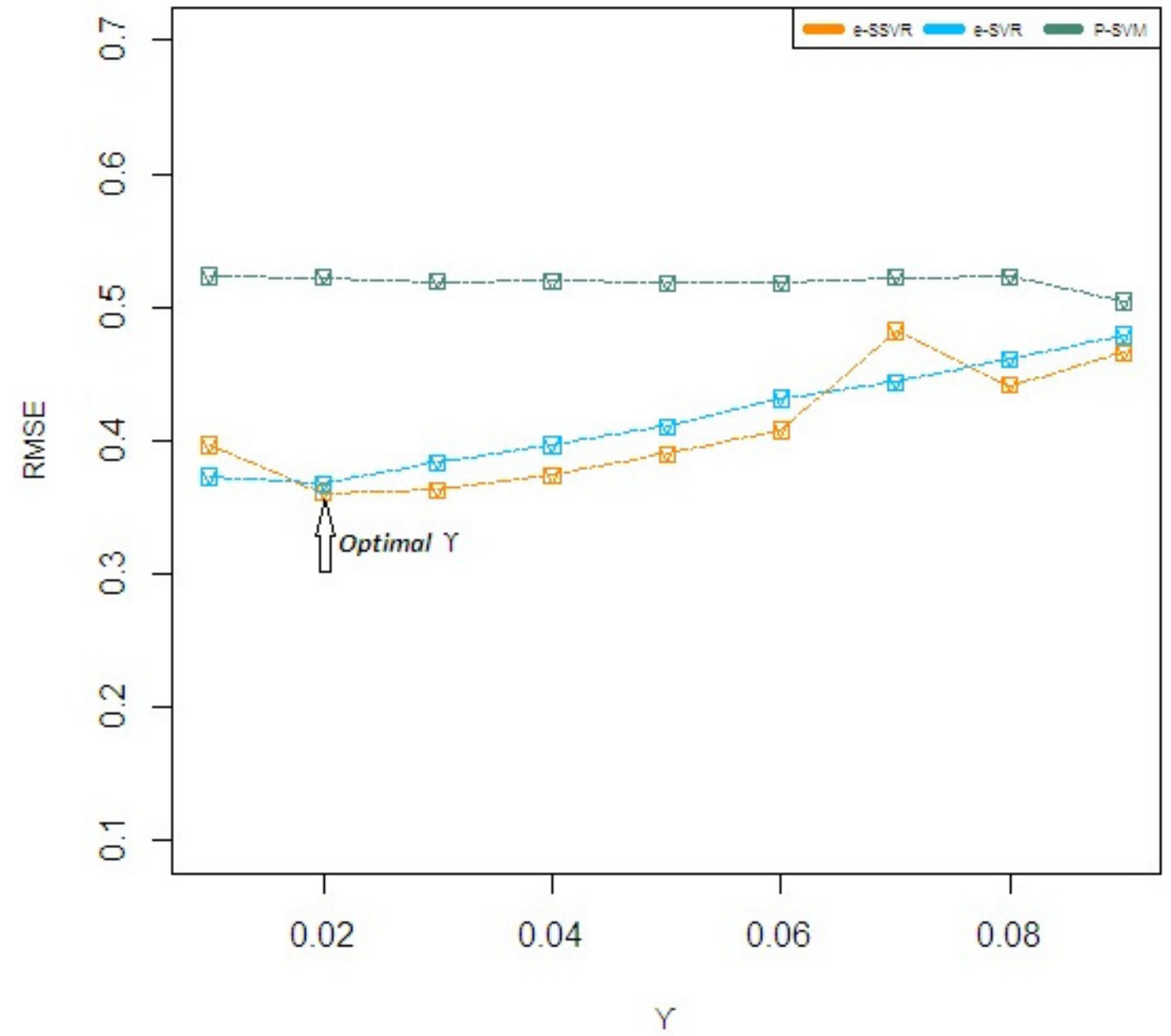

Figure 3a. The selection of the optimal ϒ(0.02) for ε-SSVR(C =1000, ε=0.1)

Parameter tuning was conducted in ε-SSVR, where the ϒparameter in the Gaussian kernel function was varied from 0.01 to 0.09 in steps 0.01 to select optimal parameter. The value of ϒ is updated based on the minimization LOO tuning error rather than directly minimizing the training error. The curves of RMSE versus the gamma(ϒ) is shown in Figure 3a and Figure 3b.

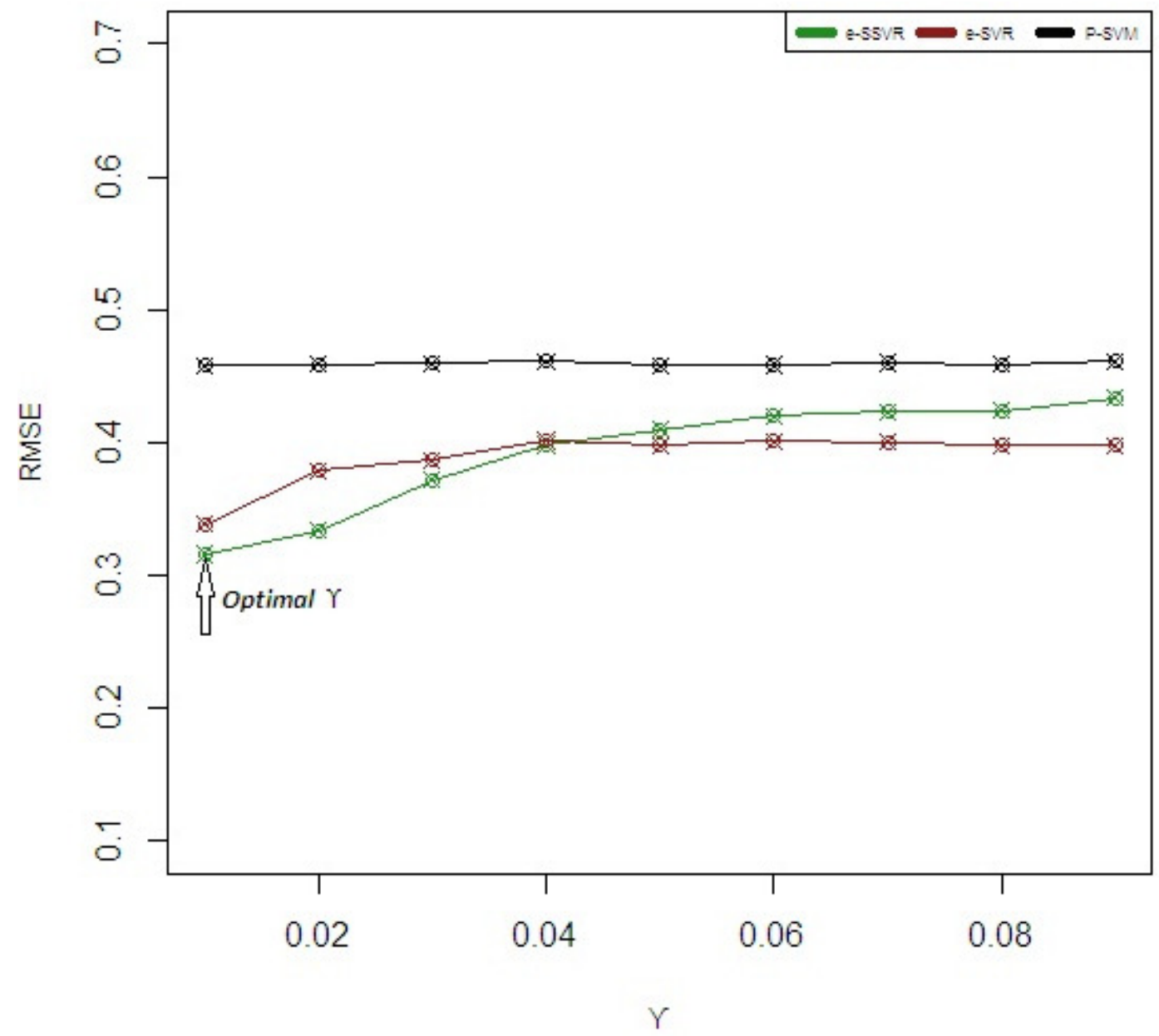

Figure 3b. The selection of the optimal ϒ(0.01) for ε-SSVR(C =1000000, ε=0.1)

For the Full descriptordataset , The RMSE value for ε-SSVR model 0.3607 is small for selected optimal parameter ϒ , compared to RMSE values for other two models i.e. LIBSVM(ε-SVR) and P-SVM are 0.3675 and 0.5224. Similarly , for the Reduced descriptor dataset , The RMSE value for ε-SSVR model 0.3161 is small for selected optimal parameterϒ, compared to RMSE values for other two models i.e.LIBSVM(ε-SVR) and P-SVM are 0.3386 and 0.4579.

The statistical parameters calculated for the ε-SSVR, LIBSVM(ε-SVR) and P-SVM models are represented in Table 2 and Table 3.

Table 2. Performance Comparison between ε-SSVR,ε-SVR and P-SVM for Full descriptor dataset

| **Algorithm** | **(ε, C,ϒ)** | **Train Error($R^2$)** | **Test Error($R^2$)** | **MAE** | **SRE** | **SEP** | **REP(%)** |
|---|---|---|---|---|---|---|---|
| ε-SSVR<br>ε-SVR<br>P-SVM | (0.1,1000,0.0217) | 0.9790<br>0.9825<br>0.8248 | 0.8183<br>0.8122<br>0.6166 | 0.0994<br>0.0918<br>0.2510 | 0.1071<br>0.0979<br>0.3093 | 0.3679<br>0.3741<br>0.5345 | 53.7758<br>54.6693<br>78.1207 |
| ε-SSVR<br>ε-SVR<br>P-SVM | (0.1,8350,0.0217) | 0.9839<br>0.9825<br>0.8248 | 0.8226<br>0.8122<br>0.6166 | 0.0900<br>0.0918<br>0.2510 | 0.0939<br>0.0979<br>0.3093 | 0.3636<br>0.3741<br>0.5345 | 53.1465<br>54.6693<br>78.1207 |
| ε-SSVR<br>ε-SVR<br>P-SVM | (0.1,1000,0.02) | 0.9778<br>0.9823<br>0.8248 | 0.8181<br>0.8113<br>0.6186 | 0.1019<br>0.0922<br>0.2506 | 0.1100<br>0.0984<br>0.3093 | 0.3681<br>0.3750<br>0.5332 | 53.8052<br>54.8121<br>77.9205 |

In these tables, statistical parameters R-square ($R^2$) ,Mean absolute error (MAE),2-N Normalization(SRE), standard error of prediction (SEP) and relative error of prediction (REP%) obtained by applying the ε-SSVR, ε-SVR and P-SVM methods to the test set indicate a good external predictability of the models.

Table 3. Performance Comparison between ε-SSVR,ε-SVR and P-SVM for Reduced descriptor dataset

| Algorithm | ( ε, C,ϒ) | Train Error($R^2$) | Test Error($R^2$) | MAE | SRE | SEP | REP(%) |
|---|---|---|---|---|---|---|---|
| ε-SSVR<br>ε-SVR<br>P-SVM | (0.1,1000000,0.02) | 0.9841<br>0.9847<br>0.8001 | 0.8441<br>0.7991<br>0.7053 | 0.0881<br>0.0827<br>0.2612 | 0.0931<br>0.0914<br>0.3304 | 0.3408<br>0.3870<br>0.4687 | 49.8084<br>56.5533<br>68.4937 |
| ε-SSVR<br>ε-SVR<br>P-SVM | (0.1,10000000,0.01) | 0.9849<br>0.9829<br>0.8002 | 0.8555<br>0.8397<br>0.7069 | 0.0851<br>0.0892<br>0.2611 | 0.0908<br>0.0967<br>0.3303 | 0.3282<br>0.3456<br>0.4673 | 47.9642<br>50.5103<br>68.3036 |
| ε-SSVR<br>ε-SVR<br>P-SVM | (0.1,1000000,0.01) | 0.9796<br>0.9829<br>0.8002 | 0.8603<br>0.8397<br>0.7069 | 0.0964<br>0.0892<br>0.2611 | 0.1056<br>0.0967<br>0.3303 | 0.3226<br>0.3456<br>0.4673 | 47.1515<br>50.5103<br>68.3036 |

An experimental results show that experiments carried out from reduced descriptor datasets shows good results rather than full descriptor dataset. As from can be seen from table 4 , the results of ε-SSVR models are better than those obtainedby ε-SVR and P-SVM models for Reduced descriptor data set.

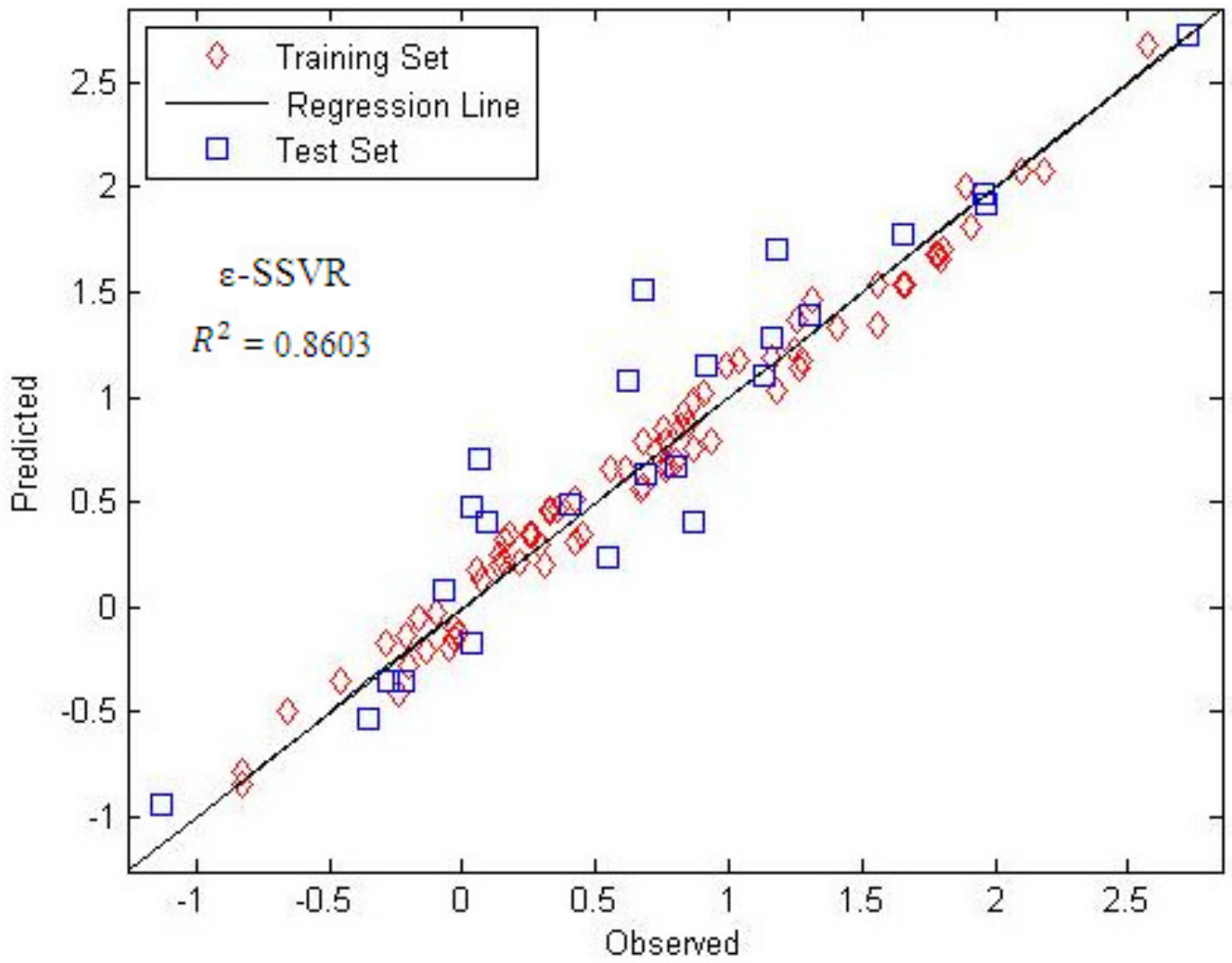

Figure 4. Correlation between observed and predicted values for training set and test set generated by ε-SSVR

Figure4, 5and 6 are the scatter plot of the three models, which shows a correlation between observed value and ant tuberculosisactivity prediction in the training and test set.

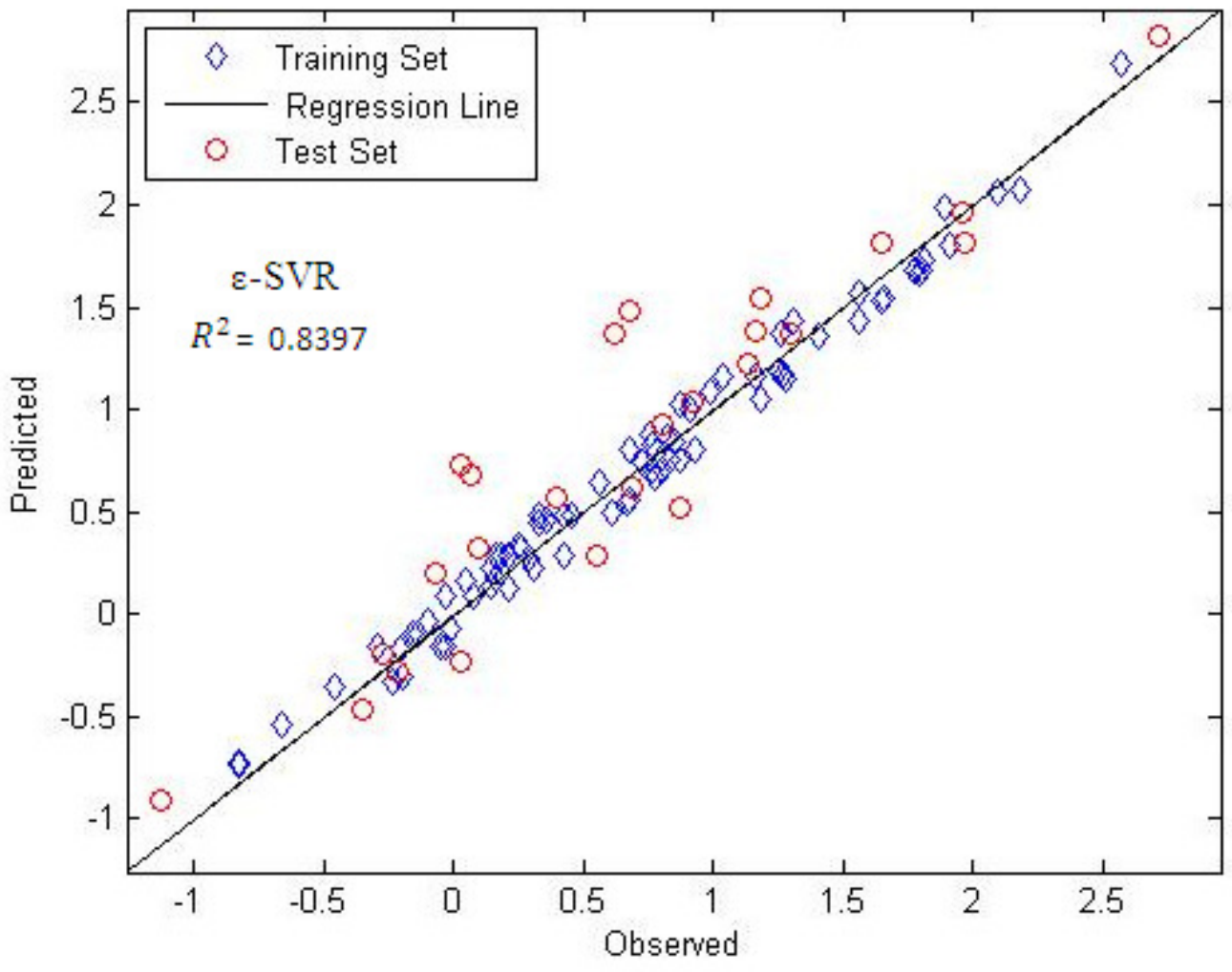

Figure 5. Correlation between observed and predicted values for training set and test set generated by ε-SVR

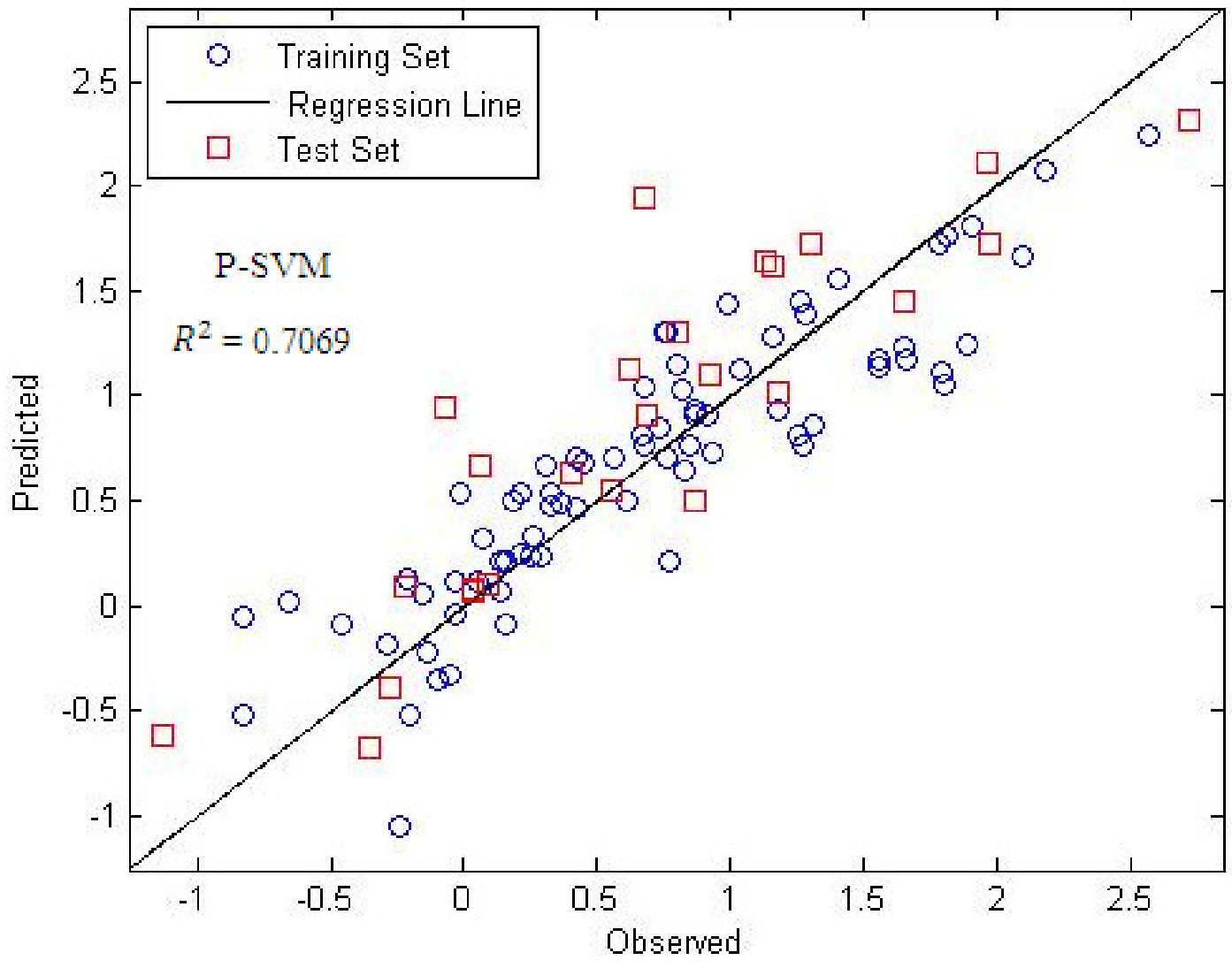

Figure 6. Correlation between observed and predicted values for training set and test set generated by P-SVM algorithm

Our numerical results have demonstrated that ε-SSVR is a powerful tool for solving regressionProblems handle the massive data sets without scarifying any prediction accuracy. In the tuning process of these experiments, we found out that LIBSVM and P-SVM become very slow when the control parameter $C$ becomes bigger, while ε-SSVR is quite robust to the control parameter $C$. Although we solved the ε-insensitive regression problem is an unconstrained minimization problem.

# 4CONCLUSION

In the present work, ε-SSVR, which is a smooth unconstrained optimization reformulation of the traditional quadratic program associated with a ε-insensitive support vector regression.We have compared the performance of, ε-SSVR, LIBSVM and P-SVM models with two datasets. The obtained results show that ε-SSVR can be used to derive statistical model with better qualities and better generalization capabilities than linear regression methods. E-SSVRalgorithm exhibits the better overall performance and a better predictive ability than the LIBSVM and P-SVM models. The experimental results indicate ε-SSVR has high precision and good generalization ability.

# ACKNOLDGEMENTS

We gratefully thank to the Department of Computer Science Mangalore University, Mangalore India for technical support of this research.

## Authors

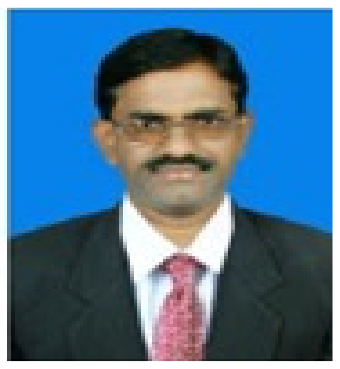

**Doreswamy**received B.Sc degree in Computer Science andM.Sc Degree inComputer Science from University of Mysore in 1993 and 1995 respectively. Ph.Ddegree in Computer Science from Mangalore University in the year 2007. Aftercompletion of his Post-Graduation Degree, he subsequently joined and served asLecturer in Computer Science at St.Joseph's College, Bangalore from 1996-1999.Then he has elevated to the position Reader in Computer Science at Mangalore Universityin year 2003. He was the Chairman of the Department of Post-Graduate Studies and researchin computer science from 2003-2005 and from 2009-2008 and served at varies capacitiesin Mangalore University at present he is the Chairman of Board of Studies and AssociateProfessor in Computer Science of Mangalore University. His areas of Researchinterestsinclude Data Mining and Knowledge Discovery,ArtificialIntelligence and Expert Systems, Bioinformatics ,Molecular modelling and simulation ,Computational Intelligence ,Nanotechnology, ImageProcessing and Pattern recognition. He has been granted a Major Research project entitled "Scientific Knowledge DiscoverySystems(SKDS) for Advanced Engineering Materials Design Applications" fromthe funding agency University Grant Commission, New Delhi, India. Hehas been published about 30 contributedpeer reviewed Papers at national/International Journal and Conferences.Hereceived SHIKSHA RATTAN PURASKAR for his outstanding achievementsin the year 2009 and RASTRIYA VIDYA SARASWATHI AWARD for outstanding achievement in chosenfield of activityin the year 2010.

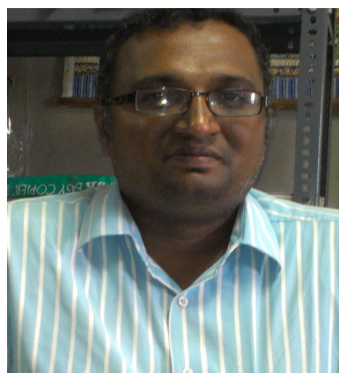

**ChanabasayyaM. Vastrad**received B.E. degree and M.Tech.degree in the year2001 and 2006 respectively. Currently working towards his Ph.D Degree in Computer Science andTechnology under the guidance of Dr. Doreswamyin the Department of Post-Graduate Studies and Research in Computer Science, Mangalore University.